\documentclass[twocolumn,floatfix,aps,eqsecnum,superscriptaddress]{revtex4-2}
\usepackage{graphicx}
\usepackage{amsmath}
\usepackage{amssymb}
\usepackage{bm}
\usepackage{hyperref}

\begin{document}

\title{Symmetric mass generation of interacting chiral fermions\\
on a one-dimensional lattice without fermion doubling}

\author{V. A. Zakharov}
\thanks{These two authors contributed equally.}
\affiliation{Instituut-Lorentz, Universiteit Leiden, P.O. Box 9506, 2300 RA Leiden, The Netherlands}

\author{Atsushi Ueda}
\thanks{These two authors contributed equally.}
\affiliation{Department of Physics and Astronomy, Ghent University, Krijgslaan 281, 9000 Gent, Belgium}

\author{Frank Verstraete}
\affiliation{Department of Physics and Astronomy, Ghent University, Krijgslaan 281, 9000 Gent, Belgium}
\affiliation{Department of Applied Mathematics and Theoretical Physics, University of Cambridge,\\ Wilberforce Road, Cambridge, CB3 0WA, United Kingdom}

\author{C. W. J. Beenakker}
\affiliation{Instituut-Lorentz, Universiteit Leiden, P.O. Box 9506, 2300 RA Leiden, The Netherlands}

\date{June 2026}

\begin{abstract}
Symmetric mass generation is the interaction-induced opening of a fermion gap without spontaneous symmetry breaking. The anomaly-free 3--4--5--0 model of Wang and Wen provides a minimal one-dimensional setting for this phenomenon, but a direct lattice realization faces two obstacles: fermion doubling for local chiral discretizations and perturbative irrelevance of the six-fermion gapping interaction. We address both obstacles. First, we formulate the model on a strictly one-dimensional tangent-fermion lattice, where a nonlocal hopping produces a single chiral branch without a mirror partner while retaining an efficient tensor-network representation. Second, we add a Hubbard-type density-density interaction (Luttinger parameter $K$) that reduces the scaling dimension of the 3--4--5--0 interaction from \(5\) to \(5K\), making it relevant for \(K<2/5\). Density-matrix renormalization group calculations show the opening of an excitation gap in this regime without the appearance of a degenerate ground state, the hallmark of symmetric mass generation.
\end{abstract}

\maketitle

\section{Introduction}
\label{intro}

Massless fermions provide an idealized framework for the exploration of strongly interacting quantum systems, both in condensed matter \cite{Giu08,Fra13,Wit16} and in particle physics \cite{Rot05,Kap09,Ton18}. In one spatial dimension (1D) they are characterized by a linear dispersion relation and a definite chirality, either left-moving or right-moving. The edge of a quantum Hall insulator provides a condensed-matter realization of 1D chiral fermions, with a quantized electrical conductance $G_{\rm H}$ and thermal conductance $G_{\rm T}$. 

The zero-mass property is protected by internal symmetries which prevent the coupling of left-movers and right-movers that would gap the spectrum. Strong interactions can generate a mass by spontaneous symmetry breaking (the Anderson-Higgs mechanism \cite{And15}), but there may be an alternative:  \textit{symmetric mass generation} (SMG) --- a process by which interactions produce mass while preserving the underlying  symmetries \cite{Eic86,Wen13,You18,Wan22,Ton22,Mou26,Has26}.

A necessary condition for SMG to work is that the fermions are anomaly-free \cite{Hoo80}. In the quantum Hall context this means that both $G_{\rm H}$ and $G_{\rm T}$ should vanish. This condition restricts the number $N_{\rm L}$ and $N_{\rm R}$ of left-movers and right-movers, and their integer charges $q_n$, since $G_{\rm H}\propto\sum_{i\in\rm{L}}q_i^2-\sum_{j\in\rm{R}}q_j^2$ and $G_{\rm T}\propto N_{\rm L}-N_{\rm R}$. 

\begin{figure}[tb]
\centerline{\includegraphics[width=0.6\linewidth]{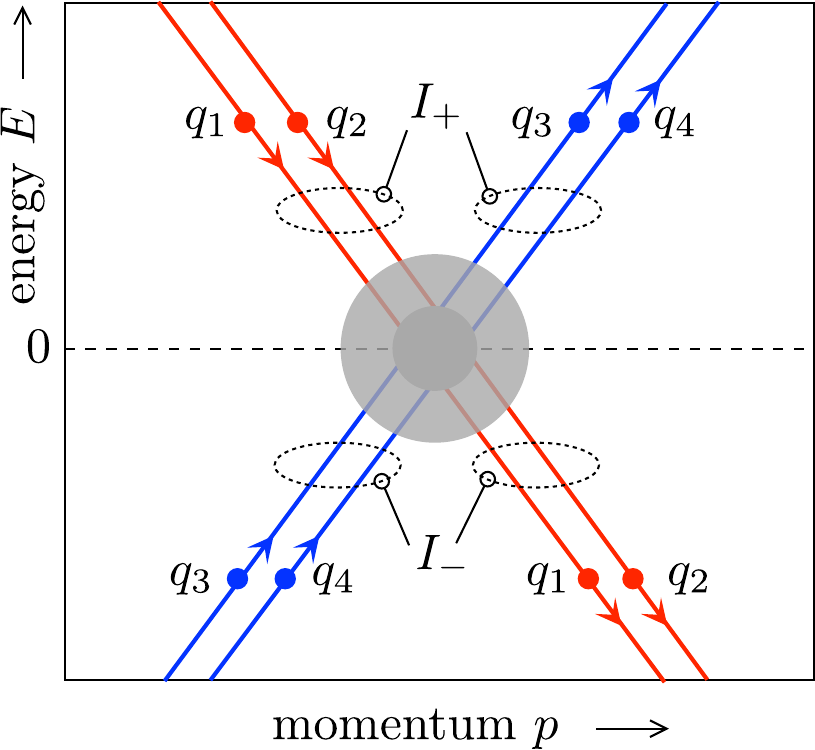}}
\caption{Dispersion of 1D chiral fermions, consisting of two branches of left-moving fermions (charges $q_1$ and $q_2$) and two branches of right-moving fermions (charges $q_3$ and $q_4$). Arrows indicate the spectral flow induced by an electric field $F$. If the negative energy branches are decoupled from the positive energy branches by a gap in the shaded energy-momentum region, the net charge current density $I_-$ that flows from negative energy into this region should vanish. One has $I_-=\sum_{n}\operatorname{sign}(v_n) q_n^2 (F/h)$, with $v_n=dE_n/dp$ the velocity in mode $n$. In the 3--4--5--0 model the gap condition $I_-=0$ is fulfilled by choosing $q_1=5$, $q_2=0$, $q_3=3$, $q_4=4$.}
\label{fig_anomaly}
\end{figure}

The 3--4--5--0 model introduced by Wang and Wen \cite{Wan19,Wan23} is a test ground for SMG of 1D chiral fermions: There are two right-movers, with charges 3 and 4, and two left-movers, with charges 5 and 0, so the model is anomaly-free ($N_{\rm L}=N_{\rm R}$ and $3^2+4^2=5^2+0^2$, see Fig.\ \ref{fig_anomaly}). The charges are co-prime, a mass term that gaps the spectrum by coupling left-movers and right-movers is prohibited by the U(1) symmetry responsible for charge conservation.

Zeng \textit{et al.} \cite{Zen22} demonstrated the interaction-induced gap opening in the 3--4--5--0 model on a 2D lattice (2+1 dimensional space-time). The additional spatial dimension was needed to accommodate the mirror fermions that appear for any local, chirality-preserving discretization of the Hamiltonian \cite{Nie81}. A fine-tuning of the hopping matrix elements decouples the mirror fermions so that their system is effectively 1D. 

Recently, related one-dimensional or exactly solvable approaches with local infinite-dimensional Hilbert spaces to anomaly-free chiral lattice gauge theories have also been proposed using emergent
translation symmetry, symmetry disentanglers, modified Villain
Hamiltonians and Euclidean models~\cite{Lu:2022xbu,Thorngren:2026abc,Seifnashri:2026abc,DeMarco19,Fazza22,Lu26,Fidkowski25,Berkowitz23, Lu26-2, Dharanikota26}.

Here we demonstrate SMG on a strictly 1D lattice. By working with Stacey's nonlocal discretization of the Hamiltonian \cite{Sta82} (with a tangent rather than a sine dispersion) we circumvent the fermion-doubling obstruction and enable the study of SMG in a minimally constrained setting. Our ``tangent fermion'' formulation builds on recent advances in quantum Monte Carlo \cite{Zak24a} and density-matrix renormalization group (DMRG) methods \cite{Hae24,Zak24b,Zak26}, that preserve computational efficiency by exploiting the fact that the nonlocal Hamiltonian corresponds to a \textit{local} generalized eigenvalue problem \cite{Pac21}.

The restriction from 2D to 1D is one way in which this study goes beyond Ref.\ \onlinecite{Zen22}. The second way is that we work around a complication of the 3--4--5--0 model, that the gapping interaction is irrelevant in the sense of the renormalization group (RG). To allow an RG scaling analysis to reliably guide the numerics, we  need to make the gapping interaction relevant. Here we achieve that by introducing an additional Hubbard-type density-density interaction, which reduces the scaling dimension of the gapping interaction and makes it perturbatively relevant for Luttinger parameter $K<2/5$. The result is a controlled setting in which the interaction responsible for SMG is the same anomaly-free 3--4--5--0 interaction, while the auxiliary Hubbard term only tunes its scaling dimension. This separates the origin of the gap from the mechanism that makes the gapping perturbation visible at accessible system sizes.

The outline of the paper is as follows. In Secs.\ \ref{sec:model} and \ref{sec_Hubbard} we present the two key ingredients of our work: We first formulate the 3--4--5--0 model on a 1D lattice without fermion doubling and then make the 3--4--5--0 interaction RG relevant by adding a Hubbard repulsion. All of this is informed by the bosonization analysis of Sec.\ \ref{sec_bosonization}. Our DMRG results are presented in Sec.\ \ref{sec_results}, focusing on the hallmark of symmetric mass generation \cite{Has26}: the opening of an excitation gap while keeping the ground state nondegenerate. We conclude in Sec.\ \ref{sec_conclude}.

\section{3--4--5--0 tangent fermions}
\label{sec:model}

The 3--4--5--0 model \cite{Wan19,Wan23} on a 1D lattice (unit lattice constant) has Hamiltonian $H=H_0+H_{3450}$, consisting of the free fermion part
\begin{align}
		H_0 ={}& \sum\limits_{n>m}t_{nm}\bigl( c^\dagger_{n,3}c^{\vphantom{\dagger}}_{m,3}+c^\dagger_{n,4}c^{\vphantom{\dagger}}_{m,4}\nonumber\\
		&-c^\dagger_{n,5}c^{\vphantom{\dagger}}_{m,5}-c^\dagger_{n,0}c^{\vphantom{\dagger}}_{m,0}\bigr)+ \text{H.c.},\label{H0def}
\end{align} 
with hopping matrix elements $t_{nm}$, and the interacting part		
\begin{align}
H_{3450}={}& \sum_n\bigl(g_1 c^{\vphantom{\dagger}}_{n,3}c^\dagger_{n,4}c^\dagger_{n+1,4}c^{\vphantom{\dagger}}_{n,5}c^{\vphantom{\dagger}}_{n,0}c^{\vphantom{\dagger}}_{n+1,0} \nonumber\\
		&+g_2 c^{\vphantom{\dagger}}_{n,3}c^{\vphantom{\dagger}}_{n+1,3} c^{\vphantom{\dagger}}_{n,4}c^\dagger_{n,5}c^\dagger_{n+1,5}c^{\vphantom{\dagger}}_{n,0}\bigr)+ \text{H.c.},\label{H3450def}
	\end{align}
with coupling constants $g_1,g_2>0$. The indices of the fermionic operators $c_{n,\alpha}$ indicate the lattice site $n\in\mathbb{Z}$ and the charge $\alpha\in\{3,4,5,0\}$. (We set the electron charge $e$ and $\hbar$ equal to unity.)

The nearest-neighbor hopping $t_{nm}=(t_0/2i)\delta_{n,m+1}$ produces the sine dispersion $E(k)=t_0\sin k$, with a spurious mirror fermion at $k=\pi$. To avoid this fermion doubling we adopt Stacey's long-range hopping \cite{Sta82}
\begin{equation}
t_{nm}=2it_0(-1)^{n-m},\label{tnmStacey}
\end{equation}
corresponding to the tangent dispersion $E(k)=2t_0\tan(k/2)$. The highly nonlocal ``all-to-all'' coupling \eqref{tnmStacey} becomes a local coupling if the Schr\"{o}dinger equation $H\psi=E\psi$ is reformulated as a generalized eigenvalue problem $P\psi=EQ\psi$, with Hermitian operators $P,Q$ that only couple nearby lattice sites \cite{Pac21}. One can thus work around the fermion-doubling obstruction without compromising computational efficiency \cite{Bee23}.

Since $\tan(k/2)$ has a positive slope in the Brillouin zone $|k|<\pi$, the fermions with charge 3 and 4 in Eq.\ \eqref{H3450def} are right-movers, while the fermions with charge 5 and 0 are left-movers. The pair of six-fermion interaction terms $\propto g_1,g_2$ conserve charge (at the origin of the U(1) symmetry),
\begin{equation}
\sum_\alpha q_\alpha \ell_\alpha=0,\label{sumqell}
\end{equation}
where $\ell_\alpha\in\mathbb{Z}$ counts how many fermion operators of charge $\alpha$ appear in the interaction term (positive $\ell_\alpha$ for a creation operator, negative $\ell_\alpha$ for an annihilation operator). 

Necessary conditions for an interaction-induced mass are that there are $N_{\rm L}=N_{\rm R}$ interaction terms, with linearly independent interaction vectors $\bm{\ell}^{(p)}$ that satisfy the null condition \cite{Hal95,Lu:2012dt,Lev13,Wan15}
\begin{equation}
\sum_{\alpha}\operatorname{sign}(v_\alpha)\ell_\alpha^{(p)}\ell_\alpha^{(p')}=0, \;\;\text{for all}\;\;p,p'.\label{vectorrule}
\end{equation}
The Hamiltonian \eqref{H3450def} has interaction vectors 
\begin{equation}
\begin{split}
&\bm{\ell}^{(1)}=(1,-2,1,2),\\
&\bm{\ell}^{(2)}=(2,1,-2,1),
\end{split}\label{elldef}
\end{equation}
for $\bm q=(3,4,5,0)$, so that both conditions \eqref{sumqell} and \eqref{vectorrule} are satisfied.

We note that the charges do not uniquely follow from the interaction vectors. For example, charges 7--11--13--1 also satisfy the charge conservation rule \eqref{sumqell}. We also note that the anomaly-free condition
\begin{equation}
\sum_{\alpha}\operatorname{sign}(v_\alpha)q_\alpha^2=0\label{anomalyfree}
\end{equation}
is not an independent condition on the charges once the interaction vectors are given --- Eq.\ \eqref{anomalyfree} follows algebraically from Eqs.\ \eqref{sumqell} and \eqref{vectorrule}.

\section{SMG at weak coupling enabled by Hubbard repulsion}
\label{sec_Hubbard}

\subsection{Renormalized scaling dimension}

We recall the basics of the scaling analysis of interacting fermions \cite{Gia03}. Interactions can open a gap in the spectrum of an infinite 1+1 dimensional system if the scaling dimension $D<2$. If $D>2$ the interactions are irrelevant, the system scales to the free-fermion limit when the size tends to infinity. 

In the non-interacting limit, the six-fermion interaction \eqref{H3450def} has scaling dimension
\begin{equation}
D_{3450}=\tfrac{1}{2}|\bm{\ell}^{(p)}|^2=5,\label{Dfree}
\end{equation}
and is therefore highly irrelevant. Strong coupling, however, may renormalize the scaling dimension, such that $H_{3450}$ becomes relevant and opens a gap in a deeply non-perturbative regime. This is the approach taken previously in Ref.\ \onlinecite{Zen22}. 

Here, we take a different route: we keep the six-fermion interaction at moderately weak coupling, where an RG scaling analysis reliably guides the numerical simulations. We add an on-site Hubbard-type density-density interaction, a four-fermion interaction that couples the density of left-movers and right-movers,
\begin{subequations}
\label{HHubbard}
\begin{align}
H_{\rm Hubbard}={}&U_{\rm H}\sum_n(\nu_3\delta\rho_{n,3}+\nu_4\delta\rho_{n,4})\nonumber\\
&\times(\nu_5\delta\rho_{n,5}+\nu_0\delta\rho_{n,0}),\\
\delta\rho_{n,\alpha}={}&c^\dagger_{n,\alpha}c^{\vphantom{\dagger}}_{n,\alpha}-\langle c^\dagger_{n,\alpha}c^{\vphantom{\dagger}}_{n,\alpha}\rangle,
\end{align}
\end{subequations}
with weight factors $\nu_\alpha$. As the Hamiltonian $H_{\rm Hubbard}$ causes no backscattering, it cannot open a gap. Nonetheless, it can modify the scaling dimension of $H_{\rm 3450}$ and thereby enable a gap opening at weak coupling.

The bosonization analysis in Sec.~\ref{sec_bosonization} shows that if we choose the weight factors according to the interaction vectors, $(\nu_3,\nu_4,\nu_5,\nu_0)=\bm{\ell}^{(p)}$, so that the Hubbard interaction takes the form
\begin{subequations}
\label{HUdef}
\begin{align}
&H_{\rm Hubbard}=\sum_{n}U_n,\\
&U_n=-U^{(1)}_{\rm H}(\delta\rho_{n,3}-2\delta\rho_{n,4})(\delta\rho_{n,5}+2\delta\rho_{n,0})\nonumber\\
&\qquad-U^{(2)}_{\rm H}(2\delta\rho_{n,3}+\delta\rho_{n,4})(-2\delta\rho_{n,5}+\delta\rho_{n,0}),
\end{align}
\end{subequations}
that with this choice of weights, the scaling dimensions $D^{(1)}_{3450},D^{(2)}_{3450}$ of the $g_1$ and $g_2$ terms in $H_{\rm 3450}$ can be {\it independently} tuned by a pair of effective Luttinger parameters $K_1,K_2$, according to
\begin{equation}
\begin{split}
&D^{(\alpha)}_{3450}=5K_\alpha,\;\;K_\alpha=\sqrt{\frac{2\pi t_0-CU^{(\alpha)}_{\rm H}}{2\pi t_0+CU^{(\alpha)}_{\rm H}}},\\
&C=\sqrt{\nu_3^2+\nu_4^2}\sqrt{\nu_5^2+\nu_0^2}=5.
\end{split}
\label{Kdef}
\end{equation}

In the following, we set $g_1=g_2$, $ U^{(1)}_{\rm H} =U^{(2)}_{\rm H}\equiv U_{\rm H}>0 $,   $K_1=K_2\equiv K\in(0,1)$. The 3--4--5--0 interaction then becomes relevant for $K<2/5\equiv K_c$. 

\subsection{Elimination of Friedel oscillations}

At nonzero Fermi wave vector $k_{\rm F}$ the electron density correlators have rapid Friedel oscillations $\propto e^{ik_{\rm F}x}$, that reduce the effectiveness of the gap opening interactions. To eliminate these we proceed as follows. 

We denote by $N_\alpha$ the expectation value of the number of fermions of charge $\alpha$ relative to a half-filled band (the free-fermion vacuum). The corresponding Fermi wave vector $k_\alpha$ is
\begin{equation}
k_\alpha = (2\pi/L)  \operatorname{sign}(v_\alpha) N_\alpha.
\end{equation}
Friedel oscillations in the correlators are absent if
\begin{equation}
\begin{split}
&\sum_\alpha \ell^{(p)}_\alpha k_\alpha = 0\Rightarrow {\cal N}^{(p)}=0,\;\;\text{for all}\;\;p,\\
&\text{with}\;\;{\cal N}^{(p)}=\sum_\alpha \operatorname{sign}(v_\alpha) \ell^{(p)}_\alpha N_\alpha.
\end{split}
\label{Friedelcondition}
\end{equation}

The Hamiltonian $H_{3450}$ conserves charge but not particle number. We can therefore not enforce $N_\alpha\equiv 0$ (half-filled band for each charge), to satisfy Eq.\ \eqref{Friedelcondition}. Instead, setting ${\cal N}^{(p)}\equiv 0$ for all $p$ is permitted because ${\cal N}^{(p)}$ is conserved by $H_{3450}$.

To see this, we note that $N_\alpha$ can change due to the interactions by an amount $\delta N_\alpha=\sum_p n^{(p)}\ell_\alpha^{(p)}$ with $n^{(p)}\in\mathbb{Z}$. The corresponding change in ${\cal N}^{(p)}$ is
\begin{equation}
\delta{\cal N}^{(p)}=\sum_{p'}n^{(p')}\sum_\alpha \operatorname{sign}(v_\alpha) \ell^{(p)}_\alpha \ell_\alpha^{(p')}=0,
\end{equation}
in view of the null condition \eqref{vectorrule}.

\section{Bosonization analysis}
\label{sec_bosonization}

\subsection{Charge rotation decouples the Luttinger liquid}

At first sight, the lattice model introduced in Sec.~\ref{sec:model} appears
rather complicated. Its interactions mix the four fermion flavors
in a nontrivial way, and it is therefore not immediately obvious
why these particular terms preserve the symmetry required for
symmetric mass generation. The structure becomes more transparent
after bosonization of the low-energy theory.

We begin with the noninteracting Hamiltonian \eqref{H0def}. The four
chiral fermions are represented by chiral bosonic fields
$\varphi_{qR/L}$, where $q$ denotes the charge and $R/L$ the
chirality (charges 3,4 right-moving, charges 5,0 left-moving). The free Hamiltonian is
\begin{align}
H_{\rm free}
&=
\frac{1}{4\pi}\int dx\,
\bigl[
(\partial_x\varphi_{3R})^2
+
(\partial_x\varphi_{4R})^2\nonumber\\
&\qquad+
(\partial_x\varphi_{5L})^2
+
(\partial_x\varphi_{0L})^2
\bigr]
\nonumber\\
&=
\frac{1}{4\pi}\int dx\,
\left[
(\partial_x\Phi_R)^2+(\partial_x\Phi_L)^2
\right],\\
\label{Hfreeboson}
\end{align}
where we have collected the chiral fields in vectors
\begin{equation}
        \Phi_R=(\varphi_{3R},\varphi_{4R}),\;\;
        \Phi_L=(\varphi_{5L},\varphi_{0L}),\;\;\Phi=(\Phi_R,\Phi_L) .
\end{equation}

The free Hamiltonian is invariant under independent orthogonal
rotations of the right-moving and left-moving fields,
\begin{equation}
        \widetilde\Phi_R=Q_R\Phi_R,\qquad
        \widetilde\Phi_L=Q_L\Phi_L .
\end{equation}
We will use this transformation to isolate neutral modes from charged modes. Similar changes of basis have been used, for example, in the study of conformal boundary conditions \cite{Yegulalp:1994eq,Smith:2019jnh,vanBeest:2023dbu}.

We take
\begin{equation}
        Q_R=
        \frac{1}{\sqrt 5}
        \begin{pmatrix}
        1 & -2\\
        2 & 1
        \end{pmatrix},
        \qquad
        Q_L=
        -\frac{1}{\sqrt 5}
        \begin{pmatrix}
        1 & 2\\
        -2 & 1
        \end{pmatrix},
\label{QRQL}
\end{equation}
chosen such that
\begin{equation}
\bm{\ell}^{(p)}=\sqrt{5}\,\bigl([Q_R]_{p,1},[Q_R]_{p,2},-[Q_L]_{p,1},-[Q_L]_{p,2}\bigr),
\end{equation}
see Eq.\ \eqref{elldef}.

With this choice,
\begin{equation}
        \bm{\ell}^{(p)}\Phi
        =
        2\sqrt 5\,\phi_p,
\label{ellPhi}
\end{equation}
where we have introduced the nonchiral two-component bosonic fields
\begin{equation}
        \phi=\tfrac{1}{2}(\widetilde\Phi_R-\widetilde\Phi_L),
        \qquad
        \theta=\tfrac{1}{2}(\widetilde\Phi_R+\widetilde\Phi_L).
\label{phitheta}
\end{equation}
The free Hamiltonian then decomposes into two decoupled Tomonaga-Luttinger liquids \cite{Kad79,Tom50,Lut63},
\begin{equation}
        H_{\rm free}
        =
        \sum_{p=1}^2
        \frac{v}{2\pi}
        \int dx\,
        \left[
        K_p(\partial_x\theta_p)^2
        +
        \frac{1}{K_p}(\partial_x\phi_p)^2
        \right],
\label{HLL}
\end{equation}
with $K_p=v=1$ at the free-fermion point.

The change of basis makes the charge symmetry explicit. A
global $\mathrm U(1)$ charge rotation acts on the
fermions as
\begin{equation}
        c_q\mapsto e^{iq\chi}c_q ,
\end{equation}
and therefore shifts the bosonic fields according to
\begin{equation}
        \Phi_\alpha\mapsto \Phi_\alpha+\chi q_\alpha .
\end{equation}
Eq.\ \eqref{QRQL} then gives
\begin{equation}
        \phi_p\mapsto \phi_p,\;\;
        \theta_1\mapsto \theta_1-\sqrt 5\,\chi,\;\;
        \theta_2\mapsto \theta_2+2\sqrt 5\,\chi .
\label{U1action}
\end{equation}
Thus the fields $\theta$ are charged, while the fields $\phi$ are neutral. 

Vertex operators built from the charged field $\theta$ are 
forbidden by charge conservation, while vertex operators built from the neutral field
$\phi$ are allowed, and can be used to introduce the 3--4--5--0 interaction. The lowest-order local vertex operators are
\begin{equation}
        V_p
        =
        \cos(\bm{\ell}^{(p)}\Phi)
        =
        \cos(2\sqrt 5\,\phi_p),
        \;\; p=1,2 .
\label{Valpha}
\end{equation}
After re-fermionization, $c_{x,q}\sim e^{i\varphi_q(x)}$, these
are precisely the two six-fermion operators in the 3--4--5--0 interaction \eqref{H3450def}.

The scaling dimension of an interaction $\propto\cos \beta\phi_p$ is $\beta^2 K_p/4$, so for $V_p$ this is $5K_p$. For
$K_p<2/5$ the cosine interaction is relevant and can pin the field $\phi_p$ to a local minimum, gapping out the fermionic excitations. When both
cosines are relevant, $K_1,K_2<2/5$, both Luttinger liquids are gapped. Since the
pinned fields are invariant under the charge rotation
\eqref{U1action}, this gap opening does not break the protecting
$\mathrm U(1)$ symmetry.

The fact that we can tune both scaling dimensions independently is a feature of the 3--4--5--0 interaction, but not necessary for SMG. In App.\ \ref{app_general} we consider the more general case. 

\subsection{Scaling dimension}

We next identify the density-density interaction that reduces the
scaling dimension of the operators $V_p$. In view of Eq.\
\eqref{HLL}, a reduction of $K_p$ is produced by the interaction
\begin{equation}
        {\cal O}_p
        =
        (\partial_x\theta_p)^2
        -
        (\partial_x\phi_p)^2
        =
        \partial_x\widetilde\varphi_{L,p}\,
        \partial_x\widetilde\varphi_{R,p} .
\label{Oalpha}
\end{equation}

Upon refermionization,
$        \partial_x\varphi_q(x)\mapsto \operatorname{sign}(v_q)\delta\rho_{x,q}$,
this interaction becomes the Hubbard-type density-density
interaction \eqref{HUdef}. It couples left-moving and
right-moving densities, but it does not backscatter the chiral
fermions and therefore does not by itself open a gap. Its role is
instead to renormalize the
free-fermion scaling dimension $D_{3450}=5$ by
\begin{equation}
        D^{(p)}_{3450}=5K_p .
\label{D3450K}
\end{equation}
The 3--4--5--0 interaction becomes relevant once
$K_p<2/5$, which is the criterion used to guide the
numerical simulations.

\subsection{Gapping without emergent ground state degeneracy}
\label{sec_gapping}

A distinctive feature of gapping by the SMG mechanism is that the ground state remains nondegenerate. Because this will play a key role in the interpretation of our numerics, we  explain it in some detail.

The four components of the chiral bosonic field $\Phi$ are defined modulo $2\pi$, the field is compactified on the four-torus $\mathbb{T}^4$. We define the map ${\cal A}$ from $\mathbb{T}^4$ to $\mathbb{T}^2$ by 
\begin{equation}
{\cal A}\Phi = A\Phi\bmod 2\pi,\;\;A=\begin{pmatrix}
        1 & -2 & 1 & 2\\
        2 & 1 & -2 & 1
        \end{pmatrix}.
\end{equation}
The two rows of $A$ are the interaction vectors $\bm{\ell}^{(1)}$ and $\bm{\ell}^{(2)}$, so that
\begin{equation}
[A\Phi]_p=\bm{\ell}^{(p)}\Phi=2\sqrt 5\,\phi_p.
\end{equation}

The map ${\cal A}$ covers the whole of $\mathbb{T}^2$ (it is surjective), because it can generate the two basis states of the two-torus:
\begin{equation}
        A (0,1,1,1)^\top=\textstyle{1\choose 0},\;\;
        A (1,1,1,0)^\top=\textstyle{0\choose 1}.
\end{equation}

Without loss of generality \cite{note3} we may choose the sign of the couplings such that the cosine potential minima are located at $\bm{\ell}^{(p)}\Phi=0$ mod $2\pi$, $p=1,2$. Its pre-image
\begin{equation}
{\cal M}=\{x\in\mathbb{T}^4:Ax=0\bmod 2\pi\}
\end{equation}
is the kernel of the map ${\cal A}$. 

The pinned ground state is nondegenerate if ${\cal M}$ is singly connected on $\mathbb{T}^4$. It is a basic theorem of algebraic geometry \cite{toric_book} that the kernel of a surjective integer-matrix map between tori has a number of connected components equal to the greatest common divisor (gcd) of the set of maximal minors of the matrix. Since $A$ is a $2\times 4$ matrix, its maximal minors are the determinants of its six $2\times 2$ submatrices. Columns 1 and 2 have gcd 5, columns 2 and 3 have gcd 3, so this already fixes the  gcd of all minors at 1 and we conclude that ${\cal M}$ has only a single connected component.

An equivalent way to state this algebraic result is that the ground state is nondegenerate because the lattice of interaction vectors,
\begin{equation}
        \Lambda=\mathrm{span}_{\mathbb Z}
        \{\bm{\ell}^{(1)},\bm{\ell}^{(2)}\}\subset \mathbb Z^4
\end{equation}
is primitive \cite{note4}. The primitivity condition means that there is no allowed local
vertex operator whose exponent is a nontrivial fractional linear
combination of the pinned fields.  In particular,
$\bm{\ell}^{(p)}/2\notin\mathbb Z^4$, so the half-harmonics
$\cos(\bm{\ell}^{(p)}\Phi/2)$ and
$\sin(\bm{\ell}^{(p)}\Phi/2)$ are not operators in the
 fermionic Hilbert space.

This is the essential difference from an ordinary symmetry-breaking
sine-Gordon problem, with a non-primitive pinning potential $\cos(2m\Phi)$. The minima
$m\Phi=0$ and $m\Phi=\pi$ are not related by the
compactification of the bosonic fields and can be distinguished by the local order
parameter $\cos(m\Phi)$.  As demonstrated for the tangent fermion Luttinger liquid in Ref.\ \onlinecite{Zak26}, this produces a degenerate
ground-state manifold, accompanied by spontaneous symmetry breaking.

\section{Results}
\label{sec_results}

We consider two related signatures of symmetric mass generation: firstly in the excitation spectrum, secondly in the occupation factor. We then show numerical DMRG results on tensor networks (see App.\ \ref{app_MPO}) that exhibit both signatures.

\subsection{Excitation gap}

\begin{figure}[tb]
\centerline{\includegraphics[width=0.8\linewidth]{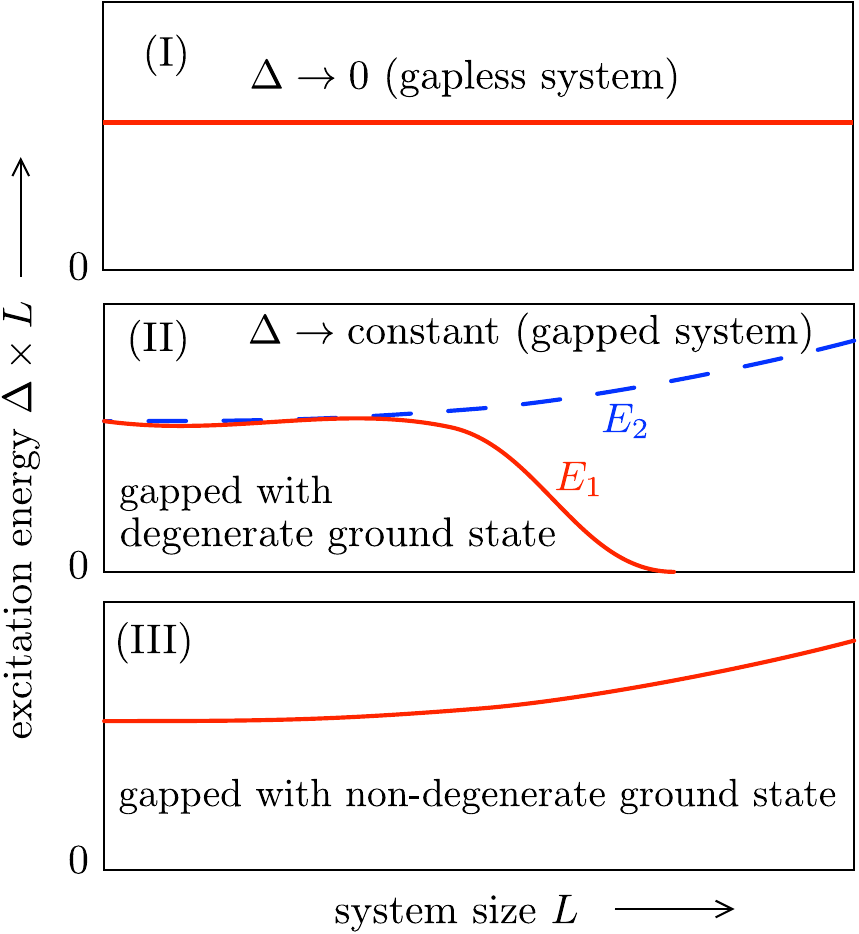}}
\caption{Comparison of the expected dependence on the system size $L$ of the excitation energy $\Delta$ (energy of an excited state relative to the ground state). Panel I shows the finite-size gap $\Delta\propto 1/L$ in a system that is gapless in the thermodynamic limit (a flat line when $\Delta\times L$ is plotted versus $L$). Panels II and III compare a gapped system with (II) or without (III) the appearance of a degenerate ground state. In case II, the symmetry is broken when $E_1$ merges with the ground state, leaving a gap to the next level $E_2$. Case III represents symmetric mass generation (SMG).}
\label{fig_scenarios}
\end{figure}

Our goal is to distinguish three types of infinite-system spectra from finite-size data: (I) a gapless system; (II) a gapped system with broken U(1) symmetry; (III) a gapped system with preserved U(1) symmetry. The presence of a finite-size gap $\Delta=\hbar v/L$ in a system of size $L$ complicates the distinction. In Fig.\ \ref{fig_scenarios} we illustrate what we expect for the $L$-dependence of the excitation gap in each of the three cases.

Case (I) is distinguished by the $1/L$ decay of the energy of the lowest excited state (relative to the ground state). In both cases (II) and (III) an excitation gap remains in the large-$L$ limit; the distinguishing feature is that in case (II) the lowest excited state $E_1$ merges with the ground state and the gap refers to the energy of the next level $E_2$. The appearance of a degenerate ground state is the signature of the Higgs mechanism for mass generation via spontaneously broken symmetry, see Ref.\ \onlinecite{Zak26} for a tangent fermion realization.

\begin{figure}[tb]
\centerline{\includegraphics[width=1\linewidth]{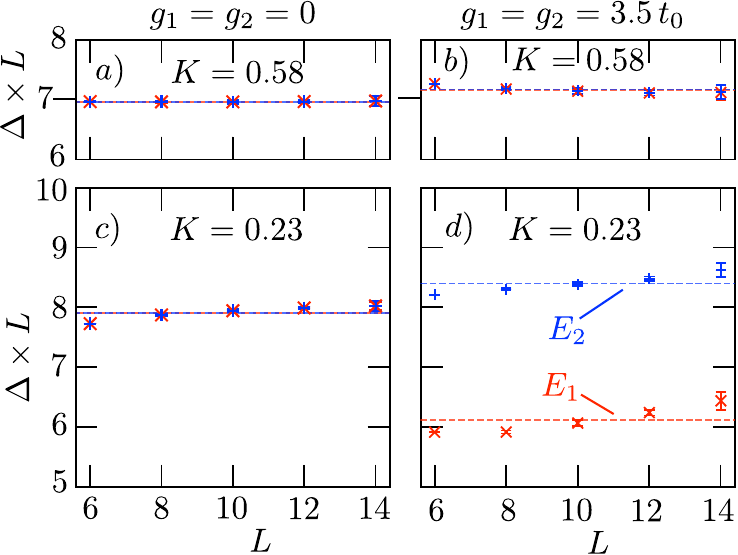}}
\caption{DMRG results for the energies of the two lowest excited states $E_1,E_2$ (measured relative to the ground state), of tangent fermions in the 3--4--5--0 model on an $L$-site 1D lattice (anti-periodic boundary conditions, bond dimension $\chi_{\rm MPS}=16384$). Data points with error bars (see App.\ \ref{app_chi}) are the numerical results, the dashed line is what we would expect for a gapless system (case I in Fig.\ \ref{fig_scenarios}). A gap opening without a degenerate ground state (case III) appears in panel \textit{d)} in the presence of both the 3--4--5--0 interaction ($g_1=g_2=3.5$) and a sufficiently strong Hubbard interaction ($K<2/5$).}
\label{fig_spectrum}
\end{figure}

The DMRG results in Fig.\ \ref{fig_spectrum}d) show the expected case III spectrum, indicative of SMG. The four panels isolate the two ingredients of the construction. 
For $g_1=g_2=0$, the Hubbard interaction changes the Luttinger parameter but does not by itself produce the SMG gap. In contrast, for $K>K_c$, the 3--4--5--0 interaction remains irrelevant, and the spectrum retains the finite-size behavior expected of a gapless system. 
Only when the 3--4--5--0 interaction is combined with a sufficiently small Luttinger parameter, $K<K_c$, do the two lowest excitation energies separate from the gapless scaling line without an accompanying collapse of $E_1$ onto the ground state. This is the finite-size signature expected for a symmetric, rather than symmetry-breaking, mass generation mechanism.

\subsection{Occupation factor}

The occupation factor $n_\alpha(k)$, the Fourier transform of the charge-$\alpha$ propagator $C_\alpha(x)=\langle c_\alpha^\dagger(x)c_\alpha(0)\rangle$, has a power-law singularity at $k=0$ for a gapless Luttinger liquid \cite{Gia03},
\begin{equation}
n_\alpha(k)\propto |k|^{[K+K^{-1}]/2-1}. 
\end{equation}
For a gapped system, instead, a smooth $k$-dependence follows from the exponential decay of the propagator.

\begin{figure}[tb]
\centerline{\includegraphics[width=1\linewidth]{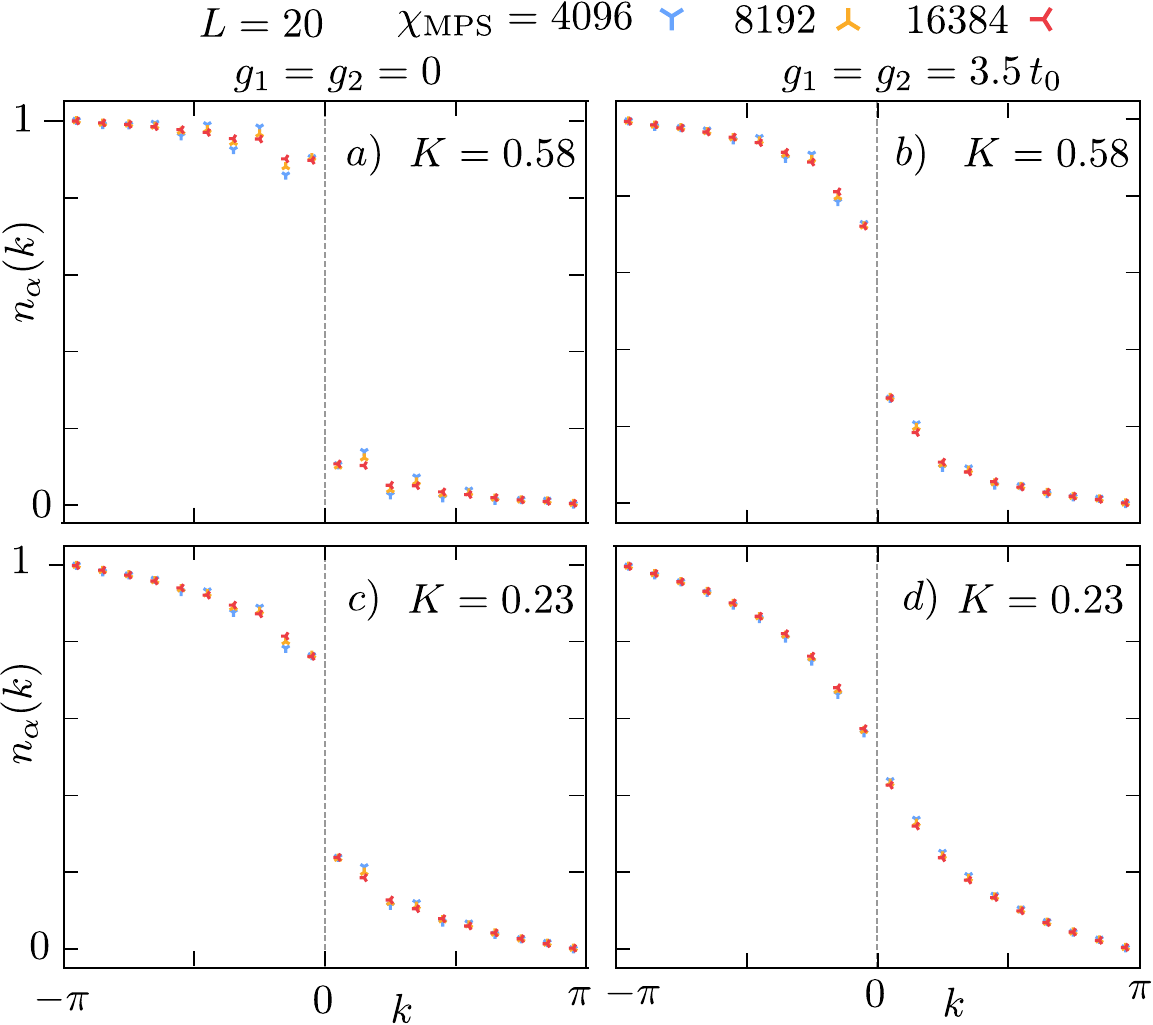}}
\caption{DMRG results for the momentum dependent occupation factor $n_\alpha(k)$ of a right-moving charge (equivalently, $\alpha=3$ or 4), computed for $L=20$ at three different MPS bond dimensions $\chi_{\rm MPS}$. The four panels correspond to the four panels in Fig.\ \ref{fig_spectrum}. The steep rise near $k=0$ is smoothed in panel \textit{d}).}
\label{fig_correlator}
\end{figure}

In a finite system, the discreteness of $k$ removes the singularity, but a steep rise remains in a gapless system, as can be seen in Fig.\ \ref{fig_correlator}, panels a,b,c). The steep rise is smoothed in panel d), consistent with the gap data from Fig.\ \ref{fig_spectrum}.

\section{Conclusion}
\label{sec_conclude}

We have studied symmetric mass generation (SMG) in the anomaly-free 3--4--5--0 model \cite{Wan19,Wan23} on a strictly one-dimensional lattice, employing fermions with a tangent dispersion (``tangent fermions'') as an alternative to lattice constructions \cite{Zen22} in which the unwanted mirror fermions are accommodated in an additional spatial dimension. To make the SMG mechanism visible in a weak-coupling scaling regime, we introduced a Hubbard-type density-density interaction that renormalizes the Luttinger parameter $K$, reducing the scaling dimension. For $K< K_c=2/5$, the gap-opening interaction becomes relevant.

The DMRG results exhibit the expected finite-size signatures of this mechanism. When either the 3--4--5--0 interaction or the sufficiently strong Hubbard renormalization is absent, the low-lying spectrum follows the behavior expected of a gapless system. When both are present, the excitation spectrum develops a gap while the ground state remains nondegenerate. The momentum occupation factor shows the corresponding smoothing of the Luttinger-liquid singularity. Taken together, these observations are consistent with interaction-induced mass generation without spontaneous breaking of the protecting \(\mathrm{U}(1)\) symmetry.

While this numerics helps to put SMG on secure ground, one may wonder whether there is a possibility for experimental support. The 3--4--5--0 interaction does not naturally appear in a physical system, but it might be realized experimentally as a synthetic interaction in a quantum simulator. What helps is that SMG is not specific for this interaction, there is a broad class of interaction vectors that enables the mechanism. We also note that the particular Hubbard interaction that we implemented to make the gapping interaction relevant can be varied with the same effect.

The tangent-fermion formalism realizes each chiral fermion flavor directly and independently on the lattice. This provides a useful framework in which the bosonization dictionary is particularly transparent. Indeed, this transparency was essential for identifying the Hubbard interaction in the simple form used here. We therefore expect tangent fermions to be useful more broadly in future studies of strongly interacting chiral fermions and their dynamics.

\acknowledgments
Figure \ref{fig_anomaly} was suggested to us by S. Polla. AU thanks L. Lootens for helpful discussions. VAZ thanks C. Xu for the explanation how SMG can be restored when the ground state has an accidental degeneracy \cite{note4}.
Our computer codes and data sets are available in a Zenodo repository at \href{https://doi.org/10.5281/zenodo.20816064}{https://doi.org/10.5281/zenodo.20816064} .\\
Research in Leiden was supported by the Netherlands Organisation for Scientific Research (NWO/OCW), as part of Quantum Limits (project number {\sc summit}.1.1016). AU was supported by BOF-GOA (Grant No.\ BOF23/GOA/021) and by FWO Junior Postdoctoral Fellowship (grant No.\ 3E0.2025.0049.01). FV acknowledges funding from the UKRI (EP/Z003342/1),  EOS (40007526) and IBOF (IBOF23/064).

\appendix

\section{Making the gapping interaction relevant when both scaling dimensions can not be tuned independently}
\label{app_general}

The interaction vectors \eqref{elldef}, and those of Ref.\ \onlinecite{note4}, share a feature that is not generic: their right-moving halves are orthogonal, $\bm{\ell}^{(1)}_{\rm R}\cdot\bm{\ell}^{(2)}_{\rm R}=0$. This allows the construction of an orthogonal transformation $Q_{\rm R}$ with its rows proportional to $\bm{\ell}^{(p)}_{\rm R}$, as in Eq.\ \eqref{QRQL}. The same condition is what makes the scaling dimensions $D^{(1)}_{3450}$ and $D^{(2)}_{3450}$ independent, so that $K_1$ and $K_2$ tune them separately, Eqs.\ \eqref{Kdef} and \eqref{D3450K}.

Not all null vectors have this feature. The pair
\begin{equation}
\bm{\ell}^{(1)}=(1,-2,1,2),\;\;\bm{\ell}^{(2)}=(1,3,-3,-1)
\label{mixedpair}
\end{equation}
satisfies the null condition \eqref{vectorrule}, but has $\bm{\ell}^{(1)}_{\rm R}\cdot\bm{\ell}^{(2)}_{\rm R}=-5$, so it does not admit decoupling into independent Luttinger liquids, each with their own interaction vector. The gapping interaction can nonetheless be made relevant. We show here how to do this for general interaction vectors.

Throughout we take $N_{\rm L}=N_{\rm R}=n$ chiral fermions and $n$ linearly independent interaction vectors obeying the null condition \eqref{vectorrule} --- the conditions under which symmetric mass generation is possible \cite{Wan23,Wan22,Hal95,Lu:2012dt,Lev13,Wan15}.

\subsection{Removing the charged fields}

As in Sec.\ \ref{sec_bosonization} we rotate the right-moving and the left-moving fields separately, $\widetilde{\Phi}_{\rm R}=Q_{\rm R}\Phi_{\rm R}$ and $\widetilde{\Phi}_{\rm L}=Q_{\rm L}\Phi_{\rm L}$, but now with arbitrary orthogonal matrices $Q_{\rm R},Q_{\rm L}$ in place of the specific choice \eqref{QRQL}. With $\bm{\phi}$ and $\bm{\theta}$ defined as in Eq.\ \eqref{phitheta}, the quadratic Hamiltonian is a set of $n$ Luttinger liquids \eqref{HLL} with parameters $K_i$, and the argument of each vertex operator splits into
\begin{subequations}
\label{gammabeta}
\begin{align}
&\bm{\ell}^{(p)}\Phi=\bm{\gamma}_p\cdot\bm{\theta}+\bm{\beta}_p\cdot\bm{\phi},\\
&\bm{\gamma}_p=Q_{\rm R}\bm{\ell}^{(p)}_{\rm R}+Q_{\rm L}\bm{\ell}^{(p)}_{\rm L},\\
&\bm{\beta}_p=Q_{\rm R}\bm{\ell}^{(p)}_{\rm R}-Q_{\rm L}\bm{\ell}^{(p)}_{\rm L}.
\end{align}
\end{subequations}
The scaling dimension of $V_p=\cos(\bm{\ell}^{(p)}\Phi)$ is then
\begin{equation}
D_p=\frac{1}{4}\sum_{i=1}^{n}\Bigl[\beta_{p,i}^2K_i+\frac{\gamma_{p,i}^2}{K_i}\Bigr],
\label{Dgeneral}
\end{equation}
and a gap opens when $D_p<2$ for every $p$.

The two contributions to Eq.\ \eqref{Dgeneral} respond oppositely to the $K_i$: in a channel where $\beta_{p,i}$ and $\gamma_{p,i}$ are both nonzero,
\begin{equation}
\beta_{p,i}^2K_i+\gamma_{p,i}^2/K_i\geq 2\bigl|\beta_{p,i}\gamma_{p,i}\bigr|
\end{equation}
whatever the value of $K_i$, so lowering $K_i$ to reduce the first term makes the second one grow and $D_p$ cannot be pushed below the criticality floor. All scaling dimensions decrease together, with no such floor, when the vertices contain the fields $\bm{\phi}$ alone,
\begin{equation}
\bm{\gamma}_p=0\;\Leftrightarrow\;Q_{\rm L}\bm{\ell}^{(p)}_{\rm L}=-Q_{\rm R}\bm{\ell}^{(p)}_{\rm R},\;\;\text{for all}\;\;p.
\label{purephi}
\end{equation}
Rotations that achieve this exist precisely because of the null condition: Eq.\ \eqref{vectorrule} states that the sets $\{\bm{\ell}^{(p)}_{\rm R}\}$ and $\{-\bm{\ell}^{(p)}_{\rm L}\}$ have the same Gram matrix,
\begin{equation}
\bm{\ell}^{(p)}_{\rm R}\cdot\bm{\ell}^{(p')}_{\rm R}=\bm{\ell}^{(p)}_{\rm L}\cdot\bm{\ell}^{(p')}_{\rm L},
\label{gram}
\end{equation}
and two sets of vectors with the same Gram matrix can be transformed into one another by an orthogonal transformation. Equation \eqref{QRQL} is the solution for the vectors \eqref{elldef}. In such a basis $\bm{\beta}_p=2Q_{\rm R}\bm{\ell}^{(p)}_{\rm R}$ and
\begin{equation}
D_p=\sum_{i=1}^{n}\bigl(Q_{\rm R}\bm{\ell}^{(p)}_{\rm R}\bigr)_i^2K_i,
\label{Dpure}
\end{equation}
which reduces to Eq.\ \eqref{Dfree} for $K_i=1$.

\subsection{Weakest interaction that opens the gap}
Equations \eqref{Dpure} and \eqref{purephi} leave both the $K_i$ and a common rotation of $Q_{\rm R}$ and $Q_{\rm L}$ free. Both choices enter Eq.\ \eqref{Dpure} only through the positive-definite matrix
\begin{equation}
G=Q_{\rm R}^{\rm T}\operatorname{diag}(K_1,K_2,\ldots,K_n)\,Q_{\rm R},
\end{equation}
whose eigenvalues are the $K_i$, in terms of which $D_p=\bm{\ell}^{(p)}_{\rm R}\cdot G\bm{\ell}^{(p)}_{\rm R}$. Since $\bm{x}\cdot G\bm{x}\geq(\min_iK_i)|\bm{x}|^2$ for every $\bm{x}$, the requirement $D_p<2$ for all $p$ implies
\begin{equation}
\min_i K_i<K_c=\frac{2}{\max_p|\bm{\ell}^{(p)}_{\rm R}|^2}=\frac{4}{\max_p|\bm{\ell}^{(p)}|^2}.
\label{Kc}
\end{equation}
The threshold is set by the longest interaction vector alone, irrespective of the basis. It gives $K_c=2/5$ for the vectors \eqref{elldef}, in agreement with Sec.\ \ref{sec_Hubbard}. For the pair \eqref{mixedpair} it gives the smaller value $K_c=1/5$, since the longer of those two vectors has $|\bm{\ell}|^2=20$.

In order to maximise the $K_i$, and therefore achieve the relevancy with the weakest Hubbard interaction, the inequality $\bm{\ell}^{(p)}_{\rm R}\cdot G\bm{\ell}^{(p)}_{\rm R}\geq(\min_iK_i)|\bm{\ell}^{(p)}_{\rm R}|^2$ needs to be saturated. It happens only on an eigenvector belonging to the smallest eigenvalue. Therefore the longest $\bm{\ell}^{(p)}_{\rm R}$ has to be an eigenvector of $G$ with the eigenvalue $K_c$. Removing this direction and repeating the procedure of maximising the $K_i$ fixes $Q_{\rm R}$ and, consequently, $Q_{\rm L}$ too.

If the $\bm{\ell}^{(p)}_{\rm R}$ are mutually orthogonal the recursion closes immediately, every interaction vector being an eigenvector of $G$, and
\begin{equation}
K_p=\frac{2}{|\bm{\ell}^{(p)}_{\rm R}|^2}=\frac{4}{|\bm{\ell}^{(p)}|^2}.
\label{Kdecoupled}
\end{equation}
Each vertex then carries a Luttinger parameter of its own, each at its own threshold: the decoupled situation of the main text is the optimal one. For the vectors \eqref{elldef} it gives $K_1=K_2=2/5$. 

For the pair \eqref{mixedpair} the recursion runs as follows. The longer vector, $\bm{\ell}^{(2)}_{\rm R}=(1,3)$ with $|\bm{\ell}^{(2)}_{\rm R}|^2=10$, has to be an eigenvector of $G$, so we align a row of $Q_{\rm R}$ with it; solving Eq.\ \eqref{purephi} for $Q_{\rm L}$ then leaves no further freedom, and
\begin{equation}
Q_{\rm R}=\frac{1}{\sqrt{10}}\begin{pmatrix}1&3\\-3&1\end{pmatrix},\;\;
Q_{\rm L}=\frac{1}{\sqrt{10}}\begin{pmatrix}3&1\\-1&3\end{pmatrix}.
\label{Qmixed}
\end{equation}
Both are orthogonal, both give $\bm{\gamma}_p=0$, and the vertex operators become
\begin{equation}
\bm{\ell}^{(1)}\Phi=-\sqrt{10}\,(\phi_1+\phi_2),\;\;
\bm{\ell}^{(2)}\Phi=2\sqrt{10}\,\phi_1 .
\end{equation}
The second cosine involves a single channel, the first involves both: the two cannot be assigned to separate Luttinger liquids. Equation \eqref{Dgeneral} gives
\begin{equation}
D_1=\tfrac{5}{2}(K_1+K_2),\;\;D_2=10\,K_1,
\end{equation}
so that relevance requires
\begin{equation}
K_1<\tfrac{1}{5},\;\;K_1+K_2<\tfrac{4}{5}.
\end{equation}
The first bound is the one that constrains the smaller parameter (in agreement with Eq.\ \eqref{Kc}), the second one then gives $K_2<3/5 - K_1$.

\section{DMRG calculation}
\label{app_MPO}

\subsection{MPO representation}

For the DMRG calculation we need to represent the Hamiltonian as a matrix-product operator (MPO) \cite{Pirvu}, a product of matrices $M^{(n)}$ that act only on site $n$. The calculation is efficient if the dimension of each matrix (the MPO bond dimension $\chi_{\rm MPO}$) is independent of the number of lattice sites $N$. Such a scale independent MPO is possible for short-range hoppings, and also for long-range hoppings that correspond to a local generalized eigenvalue problem \cite{Zak24b}.

\begin{widetext}
The free tangent fermion Hamiltonian \eqref{H0def} has MPO representation
\begin{equation}
H_{0}=[M_{0}^{(1)}M_{0}^{(2)}\cdots M_{0}^{(N)}]_{1,10}
\end{equation}
with bond-dimension 10 matrices
	\begin{align}
	M_{0}^{(n)}=
		\begin{pmatrix}
			1&c^{\vphantom{\dagger}}_{n,3}&c^{\dagger}_{n,3}&c^{\vphantom{\dagger}}_{n,4}&c^{\dagger}_{n,4}&c^{\vphantom{\dagger}}_{n,5}&c^{\dagger}_{n,5}&c^{\vphantom{\dagger}}_{n,0}&c^{\dagger}_{n,0}&U_n\\
			0&-1&0&0&0&0&0&0&0&{}2it_0c^{\dagger}_{n,3}\\
			0&0&-1&0&0&0&0&0&0&{}2it_0c^{\vphantom{\dagger}}_{n,3}\\
			0&0&0&-1&0&0&0&0&0&{}2it_0c^{\dagger}_{n,4}\\
			0&0&0&0&-1&0&0&0&0&{}2it_0c^{\vphantom{\dagger}}_{n,4}\\
			0&0&0&0&0&-1&0&0&0&-{}2it_0c^{\dagger}_{n,5}\\
			0&0&0&0&0&0&-1&0&0&-{}2it_0c^{\vphantom{\dagger}}_{n,5}\\
			0&0&0&0&0&0&0&-1&0&-{}2it_0c^{\dagger}_{n,0}\\
			0&0&0&0&0&0&0&0&-1&-{}2it_0c^{\vphantom{\dagger}}_{n,0}\\
			0&0&0&0&0&0&0&0&0&1
		\end{pmatrix}.
	\end{align}
We have also included the Hubbard interaction \eqref{HUdef}, which is purely on-site so it does not change the bond dimension. The bulk 3--4--5--0 interaction \eqref{H3450def} has bond dimension 6,
\begin{equation}
H_{3450}=[M_{3450}^{(1)}M_{3450}^{(2)}\cdots M_{3450}^{(N)}]_{1,6},
\end{equation}
	\begin{align}M_{3450}^{(n)}=
		\begin{pmatrix}
			1&c^{\vphantom{\dagger}}_{n,3}c^\dagger_{n,4}c^{\vphantom{\dagger}}_{n,5}c^{\vphantom{\dagger}}_{n,0}&c^\dagger_{n,0}c^\dagger_{n,5}c^{\vphantom{\dagger}}_{n,4}c^\dagger_{n,3}&c^{\vphantom{\dagger}}_{n,3}c^{\vphantom{\dagger}}_{n,4}c^\dagger_{n,5}c^{\vphantom{\dagger}}_{n,0}&c^\dagger_{n,0}c^{\vphantom{\dagger}}_{n,5}c^\dagger_{n,4}c^\dagger_{n,3}&0\\
			0&0&0&0&0&g_1c^\dagger_{n,4}c^{\vphantom{\dagger}}_{n,0}\\
			0&0&0&0&0&g_1c^\dagger_{n,0}c^{\vphantom{\dagger}}_{n,4}\\
			0&0&0&0&0&g_2c^{\vphantom{\dagger}}_{n,3}c^\dagger_{n,5}\\
			0&0&0&0&0&g_2c^{\vphantom{\dagger}}_{n,5}c^\dagger_{n,3}\\
			0&0&0&0&0&1
		\end{pmatrix}.
	\end{align}
\end{widetext}

The detailed construction of the Hamiltonian MPO for generic null vectors is available in our computer codes at a \href{https://doi.org/10.5281/zenodo.20816064}{Zenodo repository}.

\subsection{Error bar estimation}
\label{app_chi}

We represent the eigenstates of $H=H_0+H_{3450}$ by matrix product states (MPS, bond dimension $\chi_{\rm MPS}$) and carry out the tensor network DMRG algorithm \cite{Sch11} to variationally determine the ground state and the first few excited states. (We used the TeNPy Library \cite{tenpy} for these calculations.) We impose anti-periodic boundary conditions on the lattice of size $L$, with $L$ even to avoid the pole in the tangent dispersion at the Brillouin zone boundary.

While the matrix product representation of the Hamiltonian is exact, with small bond dimension $\chi_{\rm MPO}$, the wave function cannot be exactly represented by a matrix product of finite bond dimension $\chi_{\rm MPS}$. We therefore carry out the DMRG calculation at several values of $\chi_{\rm MPS}=2^p$, increasing in powers of two, and extrapolate to estimate the $\chi_{\rm MPS}\rightarrow\infty$ limit of the excitation energies $E_n$.

The error bars of this extrapolation are estimated as follows, based on two empirical observations:
\begin{itemize}
\item Firstly, we find that the calculated $E_n(\chi_{\rm MPS})$ decays monotonically with increasing $\chi_{\rm MPS}$. We therefore align the upper edge of the error bar with the energy at the largest available bond dimension $\chi_{\rm MPS}=2^{p_{\rm max}}$. 
\item Secondly, we find that a fit
\begin{equation}
E_n(\chi_{\rm MPS})=E_{n}+a_{n}\,\chi_{\rm MPS}^{-\alpha_n}\label{fit}
\end{equation}
to four subsequent values of $\chi_{\rm MPS}$ gives a monotonically increasing fit parameter $E_n$ as larger and larger bond dimensions are included in the fit. We therefore align the lower edge of the error bar with the fit up to $2^{p_{\rm max}}$.
\end{itemize}

\begin{figure}[tb]
\centerline{\includegraphics[width=0.8\linewidth]{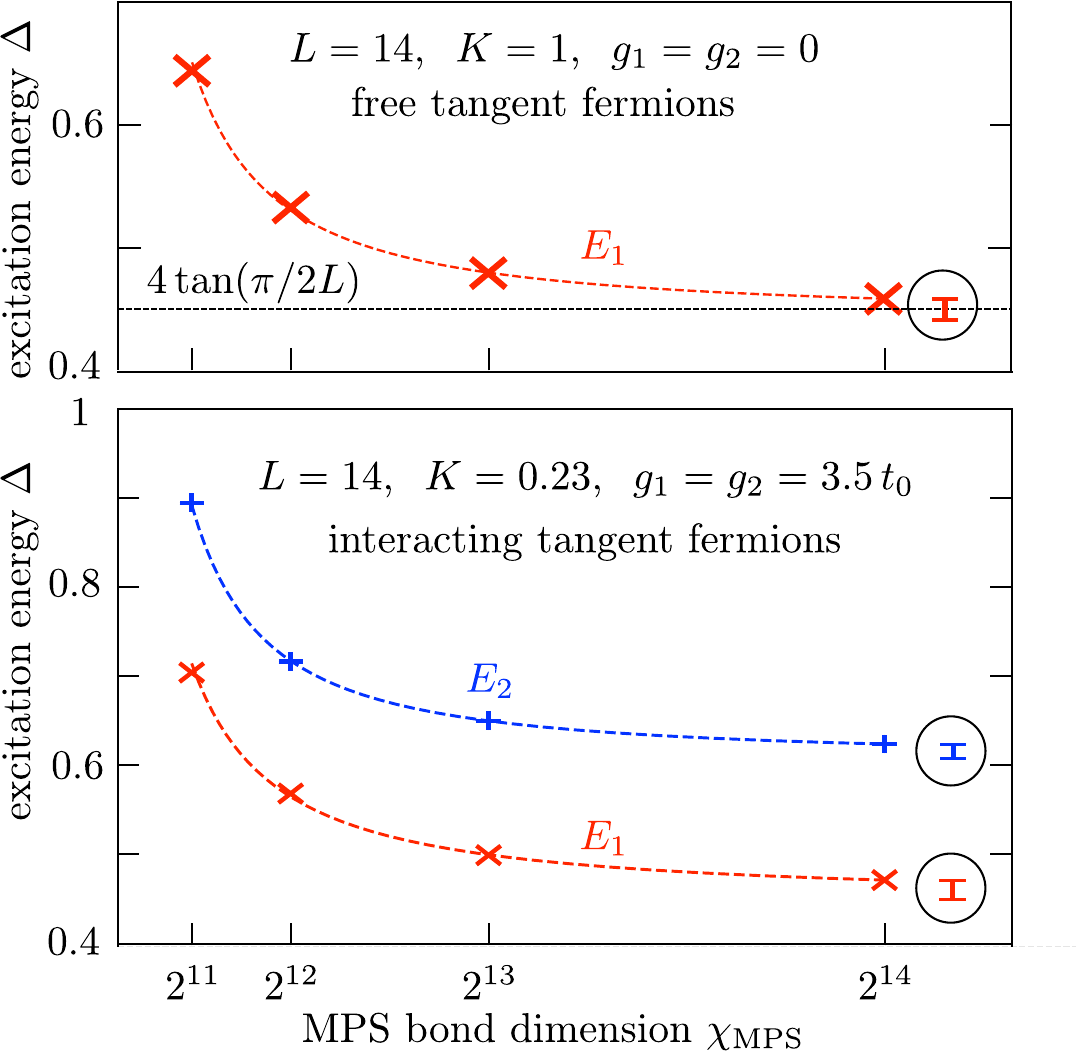}}
\caption{Dependence of the excitation energies on the MPS bond dimension $\chi_{\rm MPS}$ in the DMRG calculation. The encircled error bars of the $\chi_{\rm MPS}\rightarrow\infty$ limit are obtained by the procedure explained in the text. The upper edge of the error bar is aligned with the energy at the largest bond dimension, the lower edge of the error bar is the extrapolation of the fit \eqref{fit}. The upper panel, for the noninteracting case (when $E_1=E_2$), is a test to show that the error bars bracket the exact value $4\tan(\pi/2L)$ of the energy of the first excited state.
}
\label{fig_chiconv}
\end{figure}

Figure\ \ref{fig_chiconv} illustrates this procedure, both for the interacting case and for the case of free tangent fermions. The latter case is a test, we can compare with the known value of the excitation energy [$E_1=4\tan(\pi/2L)$, an excitation from $-2\tan(\pi/2L)$ to $+2\tan(\pi/2L)$], which is properly bracketed by the error bar. 

\begin{thebibliography}{99}
\bibitem{note3} The single-connectedness argument of Sec.\ \ref{sec_gapping} carries through if the minimum of the cosine potential is not at zero, because the translation on the torus is a continuous deformation that preserves all topological properties, including connectedness.
\bibitem{note4} As an example of a non-primitive set of interaction vectors, we note $\bm{\ell}^{(1)}=(1, 3, -3, -1)$, $\bm{\ell}^{(2)}=(-3, 1, 1, -3)$. These satisfy the conditions \eqref{sumqell} and \eqref{vectorrule}, but they do not span a primitive lattice: The fractional linear combination $\bm{L}=\tfrac{1}{2}\bm{\ell}^{(1)}+\tfrac{1}{2}\bm{\ell}^{(2)}=(-1,2,-1,-2)$ defines a local vertex operator ${\cal O}=e^{i\bm{L}\Phi}$ that can distinguish inequivalent pinned bosonic fields: If $\cos(\bm{\ell}^{(p)}\Phi)=2\pi n_p$, then ${\cal O}=(-1)^{n_1+n_2}$ takes on the value $\pm 1$ on the two inequivalent minima of the cosine potential. This is an accidental ground-state degeneracy, which can be removed by including in the Hamiltonian the operator ${\cal O}$ to favor one of the two minima without breaking ${\rm U}(1)$ symmetry, hence without compromising SMG.
\end{thebibliography}
\end{document}